\documentclass[11pt, oneside]{article}   	
\usepackage{geometry}                		
\usepackage{amsmath}								
\usepackage{amssymb}
\usepackage{subcaption}
\pdfoutput=1
\usepackage{jheppub}
\usepackage{graphicx}

\allowdisplaybreaks

\newcommand{\be}{\begin{equation}}
\newcommand{\ee}{\end{equation}}
\newcommand{\bea}{\begin{eqnarray}}
\newcommand{\eea}{\end{eqnarray}}
\newcommand{\Tad}{\mathrm{Tad}}

\def\presub{\vspace{.5cm} \noindent}
\def\bse{\begin{subequations}}
\def\ese{\end{subequations}}

\def\IZ{\relax\ifmmode\hbox{Z\kern-.4em Z}\else{Z\kern-.4em Z}\fi}
\newcommand{\non}{\nonumber \\}

\def\del{{\partial}} 
\def\presub{\vspace{.5cm} \noindent}

\def\bi{\begin{itemize}} \def\ei{\end{itemize}}

\def\({\left(} \def\){\right)}

\newcommand{\al}{\alpha}

\newcommand{\ep}{\epsilon}

\title{ Numerator Seagull and Extended Symmetries of Feynman Integrals}
\author{Barak Kol, Amit Schiller and Ruth Shir}
\emailAdd{barak.kol, amit.schiller, ruth.shir@mail.huji.ac.il }

\affiliation{Racah Institute of Physics
The Hebrew University,
Jerusalem 9190401 Israel }

\abstract{The Symmetries of Feynman Integrals (SFI) method is extended for the first time to incorporate an irreducible numerator. This is done in the context of the so-called vacuum and propagator seagull diagrams, which have 3 and 2 loops, respectively, and both have a single irreducible numerator. For this purpose, an extended version of SFI (xSFI) is developed. For the seagull diagrams with general masses, the SFI equation system is found to extend by two additional equations. The first is a recursion equation in the numerator power, which has an alternative form as a differential equation for the generating function. The second equation applies only to the propagator seagull and does not involve the numerator. We solve the equation system in two cases: over the singular locus and in a certain 3 scale sector where we obtain novel closed-form evaluations and epsilon expansions, thereby extending previous results for the numerator-free case.}

\begin{document}
\maketitle

\section{Introduction}

The evaluation of Feynman integrals belongs to the computational core of Quantum Field Theory. Yet, despite extensive experience and knowledge in this field, it is agreed that a general theory for their evaluation is still lacking.

Symmetries of Feynman Integrals (SFI) is a general method introduced in \cite{Kol:2015gsa} which associates with any given Feynman diagram a system of partial differential equations. The method uses the same variations which are used in the methods of Differential Equations \cite{Kotikov:1990kg, Kotikov:1991hm, Remiddi:1997ny, Caffo:1998du, Gehrmann:1999as} and Integration By Parts \cite{Chetyrkin:1981qh}, but distinguishes itself by associating with any diagram a natural Lie group which acts on the diagram's parameter space. 

By now, SFI was further developed and numerous diagrams have been analyzed within it \cite{Kol:2016hak,Kol:2016veg,Burda:2017tcu,Kol:2018qep,Kol:2018kga,Kol:2018ugz,Kol:2018ujm,Kol:2019lfn}. Even though these diagrams are relatively basic, the analysis achieved new results including the value of the seagull diagram (to be discussed below) in a 3 mass scale sector and the value of the kite diagram (2-loop 2-leg) throughout its singular locus.

So far, SFI addressed only Feynman integrals without numerators (also known as ``irreducible scalar products"), namely integrals involving only propagators. The goal of this paper is to study the first SFI examples of Feynman integrals with numerators and correspondingly to extend the SFI equation system and to study some of its solutions.  For some of the other works on the topic of Feynman diagrams with numerators, see \cite{Gonsalves:1983nq, Davydychev:1991va, Usyukina:1994eg, Davydychev:1995mg, Chetyrkin:1997fm, Tarasov:1997kx, Anastasiou:2000kp, Groote:2004qq, Smirnov:2004ip}.

For this purpose we shall study the propagator seagull and the vacuum seagull diagrams shown in figure \ref{fig:seagull_diagrams}, 
 which are arguably the simplest diagrams which allow for irreducible numerators. More precisely, both diagrams have a single irreducible numerator. The vacuum seagull was studied within SFI in \cite{Burda:2017tcu} and the propagator seagull in \cite{Kol:2018ujm}. 
Earlier work on these diagrams includes \cite{Davydychev:2000na, Martin:2003qz, Martin:2005qm, Martin:2016bgz, Freitas:2016zmy, Martin:2017lqn} and references within \cite{Burda:2017tcu} and \cite{Kol:2018ujm}. 
 The results of  \cite{Burda:2017tcu} were incorporated in a public computer code which relates the observable Standard Model parameters to Lagrangian parameters while accounting for some 2, 3 and even 4-loop contributions \cite{Martin:2019lqd}.  

We begin in section \ref{set up} by setting up the problem and defining the integrals under study. The general definition of irreducible numerators contains the freedom of adding squares of edge currents. We pay special attention to fixing this freedom such that the irreducible numerator has simple transformation rules under the discrete symmetries of the diagram. In the space of Schwinger parameters ($\alpha$ plane) we are able to present an expression for the general numerator seagull integral.

Section \ref{extended sfi} studies the extended equation systems. In addition to variations of the form $\del_q \cdot q $, where $q$ is a schematic notation for any energy-momentum current associated with the diagram, we allow also for variations of the form $\del_q \cdot q \, (q \cdot q)$. More generally, we allow for variations with any  degree in $q \cdot q$ (higher degree in $\del_q$ is not required for the seagull diagrams and is left for future work). This procedure produces an extension of the SFI method which we call extended SFI, or xSFI in short. The relation between numerator integrals with different numerator powers is expressed in a recursion equation. Alternatively, when passing to the generating function, it becomes a differential equation.

The equation systems are studied in section \ref{solutions} and solved both on the singular locus and in the 3 mass sector that was studied in \cite{Burda:2017tcu}, wherein we present novel expressions for numerator integrals.

Section \ref{Sec:xSFI_general} generalizes this paper's finding by presenting a relation between the extended SFI equation system and the SFI group, a relation which applies to all diagrams. 

Section \ref{summary} is a summary and discussion. Finally, appendix \ref{app: sources} details relations among the source terms which appear in the xSFI equation systems and appendix \ref{sources_feynman_integrals} details the definitions of some Feynman integrals that we use. 

\newpage
\section{Set up}\label{set up}

This section presents the integral definitions of the diagrams we will be analyzing, namely the vacuum seagull and the propagator seagull, the numerator choice, and the Schwinger parameter representation of the vacuum seagull with a general numerator power.

\subsection{Definitions}

In this paper we will consider the three-loop five-propagator vacuum diagram, which we call the vacuum seagull, and the three-loop four-propagator diagram with two external legs, which we call the propagator seagull (obtained by cutting propagator 1 of the vacuum seagull).

\begin{figure}
\centering
  \begin{subfigure}[b]{0.4\textwidth}
    \includegraphics[scale=0.33]{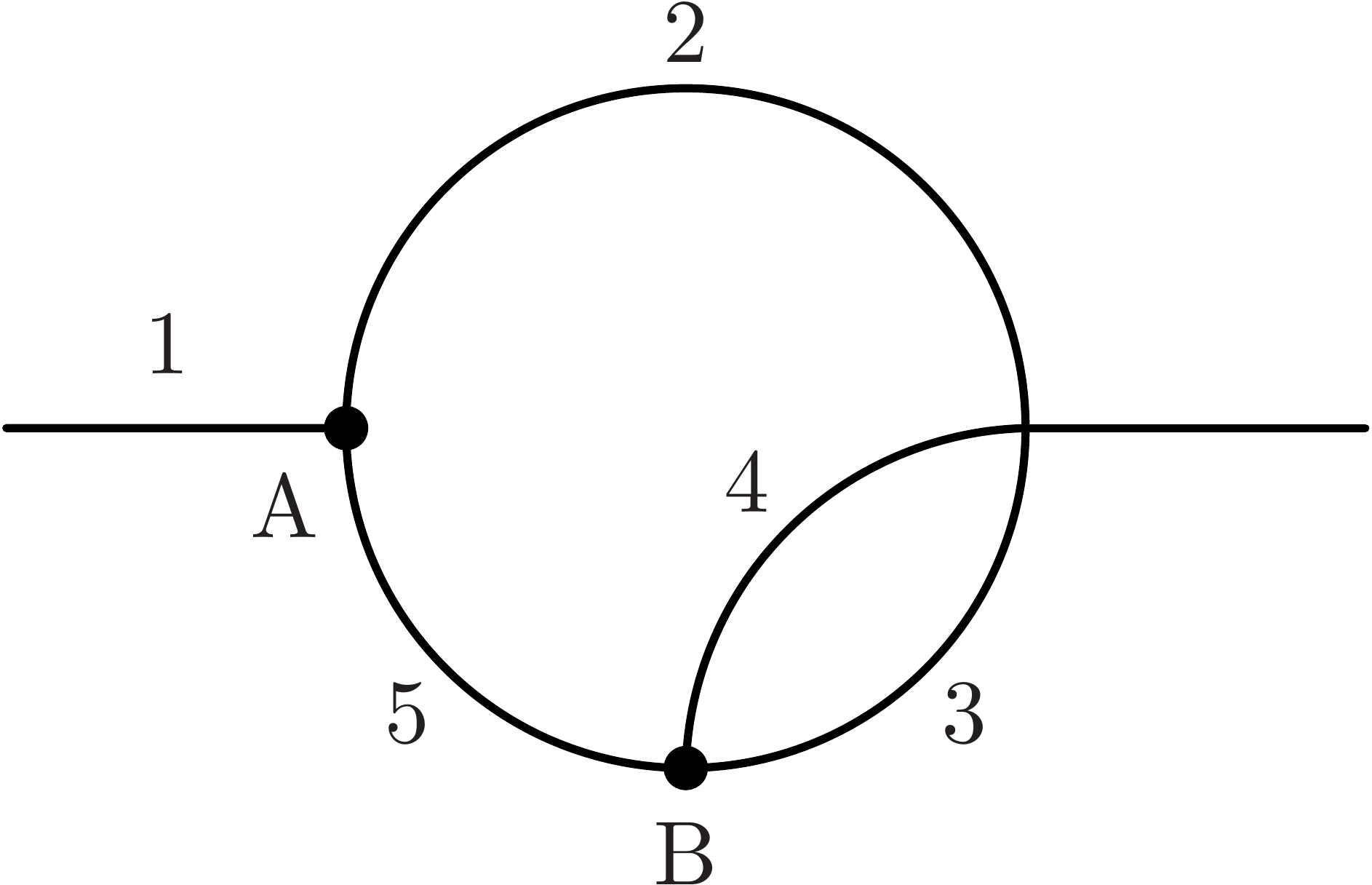}
    \subcaption{Propagator seagull}
    \label{fig:ps}
  \end{subfigure}
\hspace{0mm}
  \centering
  \begin{subfigure}[b]{0.4\textwidth}
  \centering
    \includegraphics[scale=0.33]{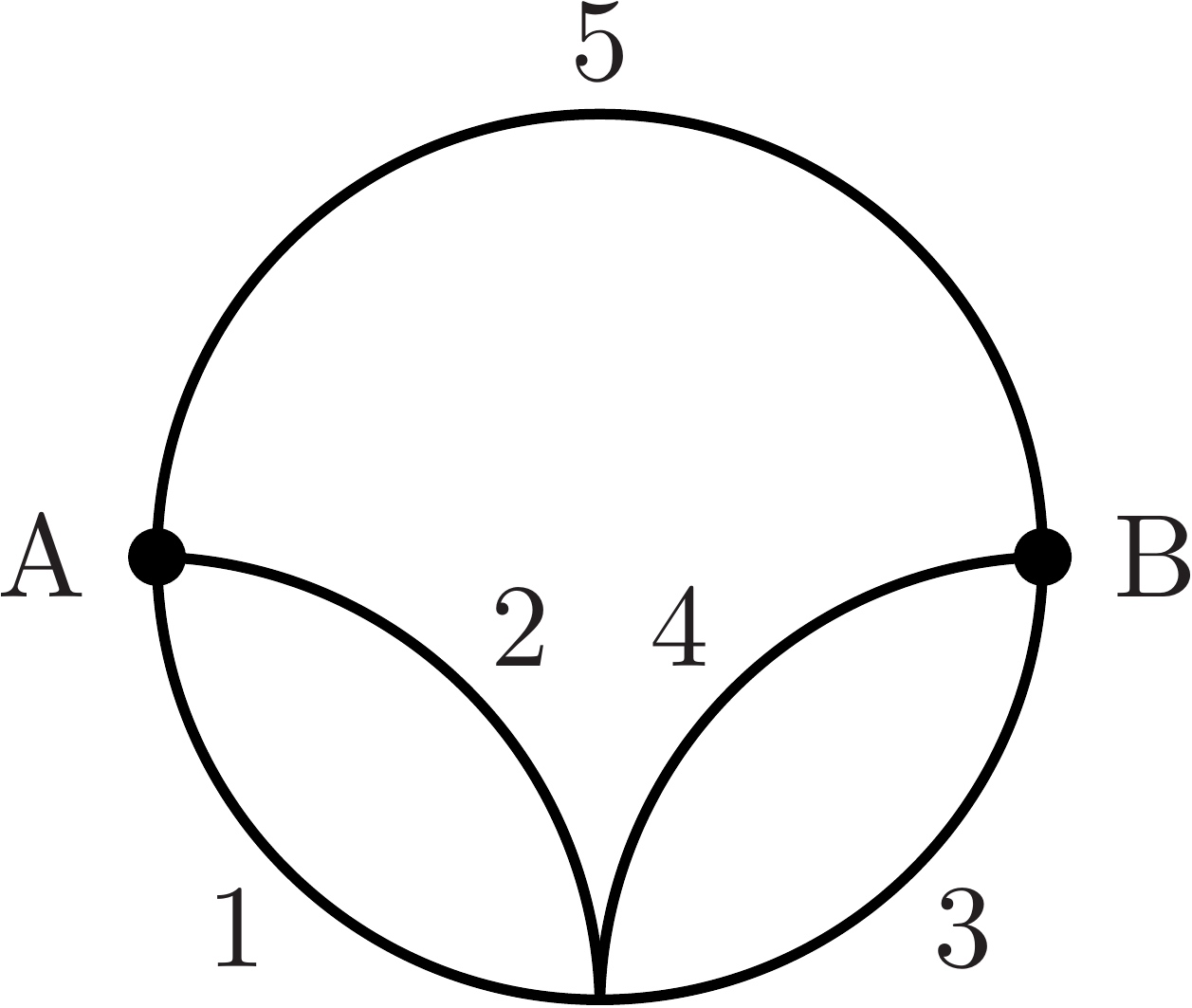}
    \vspace{5.1mm}
    \subcaption{Vacuum seagull}
    \label{fig:vs}
  \end{subfigure}
  \caption{The seagull diagrams}\label{fig:seagull_diagrams}
\end{figure}

Each diagram depends on five parameters - the vacuum seagull has five internal masses; the propagator seagull has four internal masses and one external momenta $p^\mu$. We define $x_i\equiv m_i^2$, and for the propagator seagull we also define $x_1\equiv p^2$. The numbering of the parameters is given in figure \ref{fig:seagull_diagrams}.

The propagator seagull is symmetric to exchange of propagators 3 and 4, and the vacuum seagull is symmetric to exchange of 1 and 2, 3 and 4, and to reflection. 

The propagator seagull integral is defined by
\begin{equation}\label{ps_int}
I^{PS}(x)=\int \frac{d^dl_1\,d^dl_2}{((l_1+p)^2-x_2)(l_2^2-x_3)((l_1+l_2)^2-x_4)(l_1^2-x_5)},
\end{equation}
and the vacuum seagull integral is defined by
\begin{equation}\label{vs_int}
I^{VS}(x)=\int \frac{d^dl_1\,d^dl_2\,d^dl_3}{(l_3^2-x_1)((l_1+l_3)^2-x_2)(l_2^2-x_3)((l_1+l_2)^2-x_4)(l_1^2-x_5)}.
\end{equation}
where $x$ is a collective notation for all five variables $(x) \equiv (x_1,x_2,x_3,x_4,x_5)$.
We will also denote the integrand of integral $I$ as $\tilde{I}$, e.g.
\be
I^{VS}(x)=\int d^dl_1\,d^dl_2\,d^dl_3\, \tilde{I}^{VS}(x).
\ee
When only discussing a single diagram we will drop the VS/PS denomination.

\presub {\bf Adding the numerator}. Both diagrams have 3 independent momenta (internal and external), and their space of quadratic scalars is of dimension six; as such both diagrams have one irreducible scalar product, which we will refer to as the numerator. The numerator $N$ is defined only up to adding a square. It  can be chosen in a natural way to be singlet under the discrete symmetry group of the diagram, $\Gamma$, which is guaranteed by Maschke's theorem. As such the chosen numerator is
\be
N_{VS}=(l_1+2l_2)\cdot(l_1+2l_3),
\ee
which is antisymmetric under exchanges of 1 and 2 or 3 and 4, and symmetric under reflection. The equivalent choice for the propagator seagull is 
\be
N_{PS}=(l_1+2l_2)\cdot(l_1+2p).
\ee

We can now define the propagator seagull numerator integral by
\begin{equation}\label{ps_numint}
I^{PS}_k(x)=\int \frac{d^dl_1\,d^dl_2\; N_{PS}^k}{((l_1+p)^2-x_2)(l_2^2-x_3)((l_1+l_2)^2-x_4)(l_1^2-x_5)},
\end{equation}
and the vacuum seagull numerator integral by
\begin{equation}\label{vs_numint}
I^{VS}_k(x)=\int \frac{d^dl_1\, d^dl_2\, d^dl_3\; N_{VS}^k}{(l_3^2-x_1)((l_1+l_3)^2-x_2)(l_2^2-x_3)((l_1+l_2)^2-x_4)(l_1^2-x_5)}.
\end{equation}

By a redefinition of loop currents, we can write  $I^{VS}_k$ in the following equivalent manner
\be\label{var_vs_numint}
I^{VS}_k=\int \frac{d^dl_1\,d^dl_2\,d^dl_3\; (4l_2\cdot l_3)^k}{((l_3-\frac{1}{2}l_1)^2-x_1)((l_3+\frac{1}{2}l_1)^2-x_2)((l_2-\frac{1}{2}l_1)^2-x_3)((l_2+\frac{1}{2}l_1)^2-x_4)(l_1^2-x_5)}.
\ee
This choice of currents and notation makes the discrete symmetry manifest with respect to exchanging labels, and simplifies the expression for the numerator as well.

\presub {\bf Some useful notation}.
In order to maintain the brevity of the expressions in this paper, as well as elucidate the importance of some of the quantities, we introduce some shorthand notation. 
The $s$ and $\lambda$ variables, which are associated with a 3-vertex, are defined as
\bea
s_A^i=\frac{1}{2}(x_j+x_k-x_i),\qquad i\neq j\neq k\in \{1,2,5\}\\
s_B^i=\frac{1}{2}(x_j+x_k-x_i),\qquad i\neq j\neq k\in \{3,4,5\},
\eea
and
\bea
\lambda_A=x_1^2+x_2^2+x_5^2-2(x_1 x_2+x_1 x_5 +x_2 x_5)\\
\lambda_B=x_3^2+x_4^2+x_5^2-2(x_3 x_4+x_3 x_5 +x_4 x_5),
\eea
where $A$ and $B$ denote the appropriate vertices as denoted in figure \ref{fig:seagull_diagrams}.
We also denote
\be
x_{ij}=x_i-x_j,\qquad i\neq j \in \{1,2,3,4,5\}.
\ee

\subsection{$\alpha$ plane}
In this section we will consider the vacuum seagull with numerator, and will use the choice of currents described in (\ref{var_vs_numint}).

\presub {\bf Kirchhoff-Symanzik polynomial}. In the chosen conventions we have \bea
 	\sum_{i=1}^5 \al^i\, k_i^2 &=& \al^1 (l_3-l_1/2)^2 +  \al^2 (l_3+ l_1/2)^2+ \al^3 (l_2-l_1/2)^2 +  \al^4 (l_2+ l_1/2)^2 + \al^5 (l_1)^2 = \non
 &=& A^{bc}\, l_b\, l_c 
 \eea
 where \be
    A=\left[ \begin{array}{ccc}
 	 \al^5+\(\al^1+\al^2+\al^3+\al^4\)/4	& (\al^4-\al^3)/2	 	& (\al^1-\al^2)/2 	\\
	 (\al^4-\al^3)/2			&\al^3+\al^4 &0	\\
 	(\al^1-\al^2)/2	&0  &\al^1+\al^2  \\
 \end{array} \right] ~. \ee
 Accordingly the Kirchhoff-Symanzik polynomial of the vacuum seagull is given by \be
	U = \det(A) = \al^1\, \al^2\, \al^3\, \al^4 \(\frac{1}{\al^1} + \frac{1}{\al^2}+\frac{1}{\al^3} + \frac{1}{\al^4}\) + (\al^1+\al^2)(\al^3+\al^4) \al^5
\ee

\presub {\bf Numerator integral}. The numerator integral is given by \be
	I_k := \int d\al\, dl\, N^k\, \exp\(A^{bc} l_b l_c\)\, \exp \( -\al^i \mu_i \)
\ee
where $N:=4 l_1 \cdot l_2$.

Following the dimensional recurrence relations \cite{Tarasov1996} we define \be
	I(a,\mu) = \int d\al\, dl\, \exp\(2\, a_\mu^b \,l^\mu_b\) \, \exp\(A^{bc} l_b l_c\)\, \exp \( -\al^i \mu_i \) ~.
\ee
Now \be
	I_k = \left. \( \frac{\del}{\del a^\mu_2} \frac{\del}{\del a_{\mu3}}\)^k \right|_{a=0} I(a,\mu)
\ee
while by completing the square we have \be
	I(a,\mu) = \int d\al\, \exp\( -\frac{a^b_\mu a^{c\mu} \tilde{A}_{bc}}{U}\) U^{-d/2}\, \exp \( -\al^i \mu_i \)
\ee
where $\tilde{A}$ is the adjugate matrix of $A$ (matrix of minors).

Combining the last two equations we arrive at our expression for the numerator integral in $\al$-space \be
 I_k = \int d\al\,    \left. \( \frac{\del}{\del a^\mu_2} \frac{\del}{\del a_{\mu3}}\)^k \right|_{a=0} \exp\( -\frac{a^b_\mu a^{c\mu} \tilde{A}_{bc}}{U}\) \cdot U^{-d/2}\, \exp \( -\al^i \mu_i \)
\ee
In particular we find \bea
I_1 &=& - \int d\al\,    \frac{d}{2}\, (\al^1-\al^2)(\al^3-\al^4) \frac{1}{U^{\frac{d+2}{2} }}\, \exp \( -\al^i \mu_i \) \non
I_2 &=&   \int d\al\,   4\, \left[ \( d\, \tilde{A}_{23}\)^2 +   d \( \tilde{A}_{22}  \tilde{A}_{33} +  \tilde{A}_{23}  \tilde{A}_{23}\) \right]   \frac{1}{U^{\frac{d+4}{2} }}\, \exp \( -\al^i \mu_i \)
\eea
 where in the first equality we have used $\tilde{A}_{23}=(\al^1-\al^2)(\al^3-\al^4)/4$. We note that as expected the expression for $I_1$ transforms exactly as $N$ under the graph's discrete symmetry group $\Gamma$.

\section{Extended SFI equation system}\label{extended sfi}
Here we review the SFI method, explain the idea behind extended SFI, and present the equation sets for the propagator and vacuum seagull diagrams.

\subsection{Pre-equations}
\presub {\bf SFI method}.
Before we discuss extended SFI let us first give a short review of SFI. Consider for example (\ref{ps_int}), the integral for the propagator seagull; the integral is invariant under infinitesimal linear variation of the loop momenta by any of the momenta, due to them being an integration variable. To be explicit, any transformation of the form
\be
l_i\rightarrow l_i+\epsilon_{ij}q_j,
\ee
with $i=1,2$ and $q_j={l_1,l_2,p}$, will leave (\ref{ps_int}) unchanged. It follows that the following six equations must hold:
\be \label{lvar_eq}
0=\int d^dl_1\,d^dl_2\,(\frac{\partial}{\partial l_i}\cdot q_j\tilde{I}).
\ee
Another useful equation is
\be \label{pvar_eq}
2p^2 \frac{\partial}{\partial p^2}I=\int d^dl_1\,d^dl_2\,p\cdot(\frac{\partial}{\partial p}\tilde{I}),
\ee
which is obtained from taking the variation
\be
p\rightarrow p+\epsilon_{pp}p.
\ee
As such we have an equation for each element of a group of variations
\be
\begin{pmatrix} \delta l_1 \\ \delta l_2 \\ \delta p \end{pmatrix} = T\begin{pmatrix} l_1 \\  l_2 \\ p \end{pmatrix}~.\ee 
where
\be
T \in T_{2,1} \equiv \begin{pmatrix} * & * & * \\ * & * & * \\ 0 & 0 & * \end{pmatrix}.
\ee

After some manipulation each of these equations can be brought to the form of a linear differential equation in the parameter space
\be \label{preeq_form}
F^a[I,x_i\partial_j I ,\partial_{m^2}I_1]=J^a(x),
\ee
where $i,j=1,\ldots,5$, $a=1,\ldots,7$, $F^a$ are linear functions, $\partial_{m^2}$ is a derivative by one of the internal masses squared, and the source terms $J^a$ are simpler diagrams (i.e. diagrams with one propagator omitted). 
Not all of these equations are useful if we wish to solve for $I$, since $I_1$ is an additional unknown. As such we call $\partial_{m^2} I_1$ obstructions, (\ref{preeq_form}) the pre-equations, and regard equations without any obstructions as our equation set.

The propagator seagull has four internal masses, and as such it has four obstructions ($\partial _i I_1$, $i=2,3,4,5$). It also happens that $\partial_5 I_1$ does not appear in any of the equations. As such we can get four unobstructed equations from the seven pre-equations.

Similarly for the vacuum seagull, the group of variations is $GL(3)$,
and therefore there are nine pre-equations. There are also five internal masses, and therefore five obstructions, of which $\partial_5 I^{VS}_1$ also does not appear. As such we can get five unobstructed equations for the vacuum seagull.

\presub {\bf Extended SFI pre-equations}. Equations (\ref{lvar_eq}) and (\ref{pvar_eq}) are just as true for Feynman integrals with numerators as they are for those without:
\be \label{lvar_num_eq}
0=\int d^dl_1\,d^dl_2\,(\frac{\partial}{\partial l_i}\cdot q_j \,N^k\tilde{I})=\int d^dl_1\,d^dl_2\,(N^k\frac{\partial}{\partial l_i}\cdot q_j \tilde{I}+k\,N^{k-1}\tilde{I}q_j\cdot \frac{\partial}{\partial l_i}N),
\ee
and similarly for (\ref{pvar_eq}). The resulting equation in parameter space will be of the form
\be \label{num_preeq_form}
F^a[I_k,x_i\partial_j I_k ,\partial_{m^2}I_{k+1}]+k G^a[x_i I_{k-1},I_k]=J^a_k(x),
\ee
where the indices are as in (\ref{preeq_form}), $F^a$ are the same linear functions as in  (\ref{preeq_form}), $G^a$ are additional linear functions, and the sources $J^a_k$ may include simpler integrals with numerators. So by acting with our group of variations on propagator seagull integrals with different values of $k$ we increase the number of pre-equations we have, at the price of adding additional unknown functions.

By adding to our original seven SFI pre-equations the additional seven equations achieved by acting on $I_1$, we now have fourteen equations, and as obstructions we have $I_1,\partial_i I_1$ and three $\partial_{m^2}I_2$ ($\partial_5 I_2$ does not appear in any equation), for a total of nine. So we have found a fifth equation for our original integral.

Extended SFI can also be used in order to learn about the numerator integrals; for both the vacuum and propagator integrals a linear algebraic relation can be found between $I$ and $I_1$, as there are more equations then differential expressions.

One must be careful when applying such counting arguments, for they do not assure that all the equations counted are useful - some may not be independent, and some may only contain sources. This is the case for two equations achieved from $I_k^{VS}$, meaning there are no new equations for the vacuum seagull from this extension (as expected, since we already have a number of equations equal to the number of parameters).

\subsection{Equation set}
\textbf{Propagator seagull.} The basis of variation that we have chosen in order to present the propagator seagull equations is
\be \label{ps_set var}
\begin{pmatrix}
E_1 \\ E_2 \\ E_3 \\ E_4 \\E_5
\end{pmatrix}=\begin{pmatrix}
p\frac{\partial}{\partial p}  \\ -2s_A^2\frac{\partial}{\partial l_1}l_1+(s_A^2+\frac{1}{2}x_{34})\frac{\partial}{\partial l_2}l_1-2x_5\frac{\partial}{\partial l_1}p+2s_B^4\frac{\partial}{\partial l_2}p\\\ \frac{\partial}{\partial l_2}l_2 \\\frac{\partial}{\partial l_2}(l_1 +l_2)\\\frac{\partial}{\partial l_1}l_1+\frac{\partial}{\partial l_2 }l_2 +p\frac{\partial}{\partial p} 
\end{pmatrix}I^{PS}+\begin{pmatrix}
0\\\frac{1}{2} \frac{\partial}{\partial l_2}l_1\\0\\0\\0
\end{pmatrix}I^{PS}_1,
\ee
which gives the following set of equations:
\begin{equation}\begin{split}\label{ps_set}
\begin{pmatrix} -1 \\ -2s_A^2 (d-3)\\ d-3 \\ d-3 \\2d-8    \end{pmatrix}I^{PS}-2
\begin{pmatrix} x_1 & s_A^5 & 0 & 0 & 0  \\ 0 & \lambda_A/2 & 0 & 0 &0 \\ 0 & 0 & x_3 & s_B^5 & 0 \\ 0 & 0 & s_B^5 & x_4 & 0 \\x_1 & x_2 &x_3 &x_4&x_5  \end{pmatrix}
\begin{pmatrix} \partial_1 \\ \partial_2 \\ \partial_3 \\ \partial_4 \\\partial_5 \end{pmatrix}&I^{PS}=\\=
\begin{pmatrix} -\partial_2O_5  \\ 2x_1\partial_1O_5-2s_A^5\partial_2O_5+(d-2)(O_5-O_2) \\\partial_4O_3-\partial_4O_5 \\ \partial_3O_4-\partial_3O_5\\0  \end{pmatrix}&I^{PS},
\end{split}
\end{equation}
where $O_i$ denotes an omission (or contraction) of the $i$th propagator. The sources of the second equation have been simplified using additional relations (see appendix \ref{app: sources}). The second row introduces a novel equation, beyond the standard SFI system, see \cite{Kol:2018ujm}; it was generated within xSFI and has coefficients which are higher order polynomials in $x$.

The determinant of the matrix (which we denote as $T_{PS}$) is
\be \label{detTPS}
\mathrm{det}(T_{PS})=4x_1 x_5 \lambda_A \lambda_B.
\ee
This basis has been chosen as it exhibits the symmetry and the singularities clearly, as well as having only true singularities appear as zeros of the determinant.

\presub {\bf Similarity to vacuum seagull}. Now that we have a set of five equations it is only natural to compare it to the set of the vacuum seagull. The set of the vacuum seagull can be obtained without extended SFI with the variations
\be 
\begin{pmatrix}
E_1 \\ E_2 \\ E_3 \\ E_4 \\E_5
\end{pmatrix}=\begin{pmatrix}
\frac{\partial}{\partial l_3}l_3  \\ \frac{\partial}{\partial l_3}(l_1+l_3)\\ \frac{\partial}{\partial l_2}l_2 \\\frac{\partial}{\partial l_2}(l_1 +l_2)\\\frac{\partial}{\partial l_1}l_1+\frac{\partial}{\partial l_2 }l_2 +\frac{\partial}{\partial l_3} l_3
\end{pmatrix}I^{VS},
\ee
giving the set
\begin{eqnarray}\label{vs_set}
\begin{pmatrix}  d-3 \\  d-3\\ d-3 \\ d-3 \\3d-10  \end{pmatrix}I^{VS}&-&2\begin{pmatrix}x_1 & s_A^5 & 0 & 0 & 0  \\ s_A^5 & x_2 & 0 & 0 &0 \\ 0 & 0 & x_3 & s_B^5 & 0 \\ 0 & 0 & s_B^5 & x_4 & 0 \\x_1 & x_2 &x_3 &x_4&x_5  \end{pmatrix}\begin{pmatrix} \partial_1 \\ \partial_2 \\ \partial_3 \\ \partial_4 \\\partial_5 \end{pmatrix}I^{VS}
=\begin{pmatrix}  \partial_2 O_1-\partial_2 O_5  \\ \partial_1 O_2-\partial_1 O_5 \\\partial_4O_3-\partial_4O_5 \\ \partial_3O_4-\partial_3O_5 \\0 \end{pmatrix}I^{VS}.
\end{eqnarray}
The determinant of the matrix $T_{VS}$ is
\be
\det T_{VS}=-2x_5\lambda_A\lambda_B.
\ee
As with the propagator seagull this basis was chosen in order to clearly present the symmetry and singularities.

By replacing the second equation with $2 s_A^5 Eq(1) -2 x_1 Eq(2)$ the set becomes
{\small\begin{equation}\label{vs_set_mixed}
\begin{split}
\begin{pmatrix}  d-3 \\ -2s_A^2 (d-3)\\ d-3 \\ d-3 \\3d-10   \end{pmatrix}I^{VS}-2\begin{pmatrix} x_1 & s_A^5 & 0 & 0 & 0  \\ 0 & \lambda_A/2 & 0 & 0 &0 \\ 0 & 0 & x_3 & s_B^5 & 0 \\ 0 & 0 & s_B^5 & x_4 & 0\\ x_1 & x_2 &x_3 &x_4&x_5  \end{pmatrix}\begin{pmatrix} \partial_1 \\ \partial_2 \\ \partial_3 \\ \partial_4 \\\partial_5 \end{pmatrix}&I^{VS}
=\\=\begin{pmatrix} \partial_2 O_1-\partial_2 O_5  \\ -2x_1\partial_1(O_2- O_5)+2s_A^5\partial_2(O_1-O_5) \\\partial_4O_3-\partial_4O_5 \\ \partial_3O_4-\partial_3O_5 \\0 \end{pmatrix}&I^{VS},
\end{split}
\end{equation}}
which is very similar in structure to (\ref{ps_set}). In fact the only difference in the l.h.s is that equations with $2 x_1\partial_1 I$ ($E_1$ and $E_5$) add $(d-2)I$. As for the r.h.s, if dimension equations of the simpler diagrams are used to remove all derivatives by $x_1$ the only difference will be $O_1$ terms; this is easy to see for all but $E_2$, for which one needs an additional relation in appendix \ref{app: sources}.
 Because of this equivalence any result obtained for either diagram using $E_2,E_3,E_4$, and the combination $E_1-E_5$ must be true for the other diagram up to $O_1$ terms. In fact any extended SFI equation that does not include derivatives by $x_1$ will only differ between diagrams in sources.
 
Note that even though the $T_{PS}$ and $T_{VS}$ are now identical, the singular behaviour of the set as a whole is not, as $x_1=0$ is not a singularity of (\ref{vs_set_mixed}).

\presub {\bf Higher numerator orders.}
By acting with the same variations in (\ref{ps_set var}) on $I_k^{PS}$  instead of acting on $I^{PS}$  we can get a more general version of (\ref{ps_set})
{\small \begin{eqnarray} \label{ps_set k}
\begin{pmatrix} -1+k \\ -2s_A^2 (d-3+k)\\ d-3+k \\ d-3+k \\2d-8+2k    \end{pmatrix}I^{PS}_k+k\begin{pmatrix} x_{34} \\ x_{34}(3x_1+x_2-x_5)\\ x_{12} \\ -x_{12} \\0    \end{pmatrix}I^{PS}_{k-1}-2
\begin{pmatrix} x_1 & s_A^5 & 0 & 0 & 0  \\ 0 & \lambda_A/2 & 0 & 0 &0 \\ 0 & 0 & x_3 & s_B^5 & 0 \\ 0 & 0 & s_B^5 & x_4 & 0 \\x_1 & x_2 &x_3 &x_4&x_5  \end{pmatrix}
\begin{pmatrix} \partial_1 \\ \partial_2 \\ \partial_3 \\ \partial_4 \\\partial_5 \end{pmatrix}&I^{PS}_k& = \non = \begin{pmatrix} -\partial_2O_5  \\ 2x_1\partial_1 O_5-2s_A^5\partial_2O_5+(d-2)(O_5-O_2)-kO_2\\\partial_4(O_3-O_5) \\ \partial_3(O_4-O_5)\\0  \end{pmatrix}&I^{PS}_k&+ \non
-k\begin{pmatrix} O_3-O_4  \\x_{34}(2O_5-3O_2)+(2(x_2+s_A^5)+O_2)(O_3-O_4)  \\-O_2 \\ O_2\\0  \end{pmatrix}&I^{PS}_{k-1}& \non 
\end{eqnarray}}

Doing the same for $I^{VS}_k$ we get a generalized version of (\ref{vs_set_mixed}) as well
{\small\begin{eqnarray}
\begin{pmatrix} d-3+k \\ -2s_A^2 (d-3+k)\\ d-3+k \\ d-3+k \\3d-10+2k    \end{pmatrix}I_k^{VS}+k\begin{pmatrix} x_{34} \\ x_{34}(3x_1+x_2-x_5)\\ x_{12} \\ -x_{12} \\0    \end{pmatrix}I_{k-1}^{VS}-2
\begin{pmatrix} x_1 & s_A^5 & 0 & 0 & 0  \\ 0 & \lambda_A/2 & 0 & 0 &0 \\ 0 & 0 & x_3 & s_B^5 & 0 \\ 0 & 0 & s_B^5 & x_4 & 0 \\x_1 & x_2 &x_3 &x_4&x_5  \end{pmatrix}
\begin{pmatrix} \partial_1 \\ \partial_2 \\ \partial_3 \\ \partial_4 \\\partial_5 \end{pmatrix}&I^{VS}_k&= \non
=
\begin{pmatrix} \partial_2(O_1-O_5)  \\2(x_1\partial_1-s_A^5\partial_2)(O_5-O_2-O_1)\\\partial_4(O_3-O_5) \\ \partial_3(O_4-O_5)\\0  \end{pmatrix}I^{VS}_k-k\begin{pmatrix} O_3-O_4  \\ 2(x_1+s_A^5)(O_3-O_4) \\O_1-O_2 \\ -O_1+O_2\\0  \end{pmatrix}&I_{k-1}^{VS}&~.\non
\end{eqnarray}}


\subsection{Equation for numerator diagrams}

Equation set \eqref{ps_set} was derived by pre-equation combinations with no numerators. Now, we shall consider numerators and will obtain a new equation. 

While most pre-equations yield a differential equation for $I_1$ (in terms of $I$ and sources), it is possible to select a variation which yields an algebraic equation for it. In particular, this equation will not have derivatives by $x_1$, and therefore will be true for both the propagator and seagull (up to sources). 
The variation that gives this relation for propagator seagull is
\begin{equation}
E_{num}^{PS}=\frac{\partial}{\partial l_2} \Big( \frac{1}{2} \big( x_{12}(3x_3+x_4-x_5)+\lambda_B \big)l_1-x_{12}\,x_{34}\,l_2+\lambda_B\,p \Big) I^{PS}+\frac{\partial}{\partial l_2} \Big(s_B^4l_1+x_5\,l_2  \Big) I_1^{PS},
\end{equation}
while for the vacuum seagull we replace $p$ with $l_3$.

The resulting equations are 
\be \label{numerator equation}
x_5(d-2)I_{1}=(d-2)x_{12}\,x_{34}I-J_1^{PS/VS}, 
\ee
with the source terms, after simplification (see appendix \ref{app: sources}), being
\begin{eqnarray}\label{num source}
J_1^{PS}\equiv J_1[I^{PS}]&=&(x_{34}(8-3d+4x_{2}\partial_{2}+2x_{3}\partial_{3}+2x_{4}\partial_{4})-x_{12}(2x_{3}\partial_{3}-2x_{4}\partial_{4}))O_{5}I^{PS}\non & &-(d-2)x_{12}(O_{3}-O_{4})I^{PS}+(x_{34}+\frac{\lambda_{B}}{2}(\partial_{3}-\partial_{4}))O_{2}I^{PS}\non
& &+(s_B^4\partial_{3}-s_B^3\partial_{4})(O_{5}-O_{3}-O_{4})O_{2}I^{PS},
\end{eqnarray} for the propagator seagull, and
\be\label{VS num source}
J^{VS}_{1}=J_1[I^{VS}] -\frac{1}{2}(2x_{34}+\lambda_B(\partial_3 - \partial_4))O_1 I^{VS}-(s_B^4\partial_{3}-s_B^3\partial_{4})(O_{5}-O_{3}-O_{4})O_1I^{VS}, 
\ee
for the vacuum seagull.

As we will see in section \ref{singular loci}, this equation reduces to (\ref{x5 locus}) on the locus $x_5=0$.

\subsection{Recursion relation}

We can generalize (\ref{numerator equation}) into a recursion relation by acting on integrals of general numerator order.
The variation will be
\begin{equation}
E_{num}^k=\frac{\partial}{\partial l_2} \Big( \frac{1}{2} \big( x_{12}(3x_3+x_4-x_5)+\lambda_B \big)l_1-x_{12}\,x_{34}\,l_2+\lambda_B\,p \Big) I_{k-1}^{PS}+\frac{\partial}{\partial l_2} \Big(s_B^4l_1+x_5\,l_2  \Big) I_k^{PS},
\end{equation}
where again we replace $p$ with $l_3$ for the vacuum seagull. The recursion relation obtained is
\be \label{recursion equation}
(k+d-3)x_5 I_{k}-(2k+d-4)x_{12}\,x_{34}I_{k-1}+(k-1)B(x)I_{k-2}=-J_{k}^{PS/VS},
\ee
where
\bea
B(x)&=&-x_{5}^{3}+2\left(x_{1}+x_{2}+x_{3}+x_{4}\right)x_{5}^{2}\non
& & -\left((x_{3}-x_{4})^{2}+\left(x_{1}-x_{2}\right){}^{2}+4(x_{3}+x_{4})\left(x_{1}+x_{2}\right)\right)x_{5}\nonumber\\
& & +2(x_{3}-x_{4})^{2}\left(x_{1}+x_{2}\right)+2(x_{3}+x_{4})\left(x_{1}-x_{2}\right){}^{2}]
\eea
is the Baikov polynomial\footnote{See \cite{Baikov:1996cd} for the original presentation of the Baikov polynomial.} of both the propagator and vacuum seagulls, and with the sources being
\bea \label{seagull_w_numerator_source}
J_{k}^{PS} =& &J[I_k^{PS}]=(s_{B}^{4}\partial_{3}+s_{B}^{3}\partial_{4})(O_{5}-O_{3}-O_{4})I^{PS}_{k} \non
& & +\frac{1}{2}(x_{12}\left((3x_{3}+x_{4}-x_{5})\partial_{3}-(x_{3}+3x_{4}-x_{5})\partial_{4}\right)\left(O_{3}+O_{4}-O_{5}\right) \non
& & +\left(2k\,x_{34}+\lambda_{B}(\partial_{3}-\partial_{4})\right)O_{2})I^{PS}_{k-1} \non
& & +(k-1)((2\lambda_{B}-x_{12}(2x_3+2x_4-x_5))O_{2}-\lambda_{B}O_{5})I^{PS}_{k-2}~,
\eea
for the propagator seagull, and 
\be \label{vs_w_numerator_source}
J^{VS}_{k}=J[I_k^{VS}] -\frac{1}{2}(2 k x_{34}+\lambda_B(\partial_3 - \partial_4))O_1 I^{VS}_{k-1}+(k-1)(2\lambda_B + x_{12}(2x_3+2x_4-x_5))O_1 I^{VS}_{k-2}~,
\ee
for the vacuum seagull.
It should be noted that, while it does not seem so, this source term is invariant under reflections as it should be. 

Equation \eqref{recursion equation} is one of our main results. It provides a recursion relation for numerator integrals with different powers.

\presub {\bf Generating function of numerator integrals}.

The numerator integral $I_k$ is labelled by an integer $k$. Such data can be transformed into a generating function
\be
G(z)=\sum_{k=0}^{\infty}I_{k}z^{k}
\label{def:Gz}
\ee
thereby transforming $k$ into the formal parameter $z$.

We can now transform the recursion equation \eqref{recursion equation} into the following differential equation
for $G(z)$
\bea \label{recursion diff form}
\left( x_5 \, z -2\, x_{12}\, x_{34}\, z^2 + B(x)\, z^3 \right) G'&+& \left((d-3)\, x_5-(d-2)\, x_{12}\, x_{34}\, z + B(x)\, z^2 \right)G = \nonumber \\
 &=& (d-3)\, x_5\, I-\sum_{k=1}^{\infty}J_{k}\, z^k.
\eea
This differential equation for the generating function holds for both the propagator and the vacuum seagulls, accounting for their respective sources $J_k^{PS/VS}$.

Eq. \eqref{def:Gz} defines the so-called ordinary generating function. We note that had we chosen to define the exponential generating function, we would have gotten a third order differential equation (1st order in powers of $z$).


\section{Solutions}\label{solutions}
In this section we present solutions on the singular loci as well as a closed form solution for a certain mass sector for the vacuum seagull diagram with numerator.

\subsection{Singular loci} \label{singular loci}
The singular loci are hypersurfaces in parameter space on which the equation set is singular.  On such a singular locus the equation set includes an algebraic equation and there is no need to solve the differential set of equations (see also \cite{Kalmykov:2011yy} for other algebraic relations between master integrals).  For the propagator seagull, the SFI equation set (\ref{ps_set}) is singular when $\det (T_{PS})$ given in (\ref{detTPS}) is zero, namely
\bea
x_1&=&0,\\
x_5&=&0,\\
\lambda_A&=&0,\\
\lambda_B&=&0.
\eea
These singularities are shared with the general numerator power equation set \eqref{ps_set k}, and as such on these loci additional recursion relations can be found.

We will examine each of these hypersurfaces separately, using the method of maximal minors \cite{Kol:2018qep}.
To find solutions on the loci we compute the adjunct matrix of $T$ given by   
\be
M^i_a(x)=\epsilon_{a a_1 a_2 a_3 a_4}\epsilon^{i i_1 i_2 i_3 i_4}\,T^{a_1}_{i_1}\,T^{a_2}_{i_2}\,T^{a_3}_{i_3}\,T^{a_4}_{i_4}.
\ee
On the loci hypersurfaces, $M^i_a(x)$ will factorize,
\be
M^i_a(x)\Big|_{\text{loci hypersurface}} =s(x) k_a(x) n^i(x)
\ee
where $s(x)$ is a common (scalar) factor, $k_a(x)$ is a row vector which is perpendicular to the rows of $T$ on the hypersurface, namely
\be
k_a(x)\cdot T^a_i\Big|_{\text{loci hypersurface}}=0
\ee
and $n^i(x)$ is a column vector which should be perpendicular to the columns of $T$ on the hypersurface
\be
T^a_i \cdot n^i(x)\Big|_{\text{loci hypersurface}}=0.
\ee
In other words, $k_a(x)$ defines a stabilizing generator at $x$ and $n^i(x)$ annihilates the $G$-orbit tangent space. 

The solution on the locus is then given by
\be \label{sing_loc_sol}
I(x)\Big|_{\text{loci hypersurface}}=\frac{k(x)\cdot J}{k(x)\cdot c}\Big|_{\text{loci hypersurface}}~.
\ee

\subsubsection{$x_1=0$}

For this locus we will show the calculation in greater detail as an example of use.

The adjunct matrix of $T_{PS}\Big|_{x_1=0}$ is 
\be
M^i_a(x)\Big|_{x_1=0}=-2x_5 x_{25} \lambda_B\begin{pmatrix}x_{25} &-1 &0 &0 &0 \\ 0 & 0 & 0 &0 & 0 \\0 & 0 & 0& 0 & 0 \\ 0& 0& 0& 0 &0 \\ 0& 0& 0& 0 &0 \end{pmatrix}
\ee
and decomposes as follows
\be
k_a(x)=(x_{25},-1,0,0,0)
\ee
\be
n(x)=(1,0,0,0,0).
\ee
The combination of equations defined by $k_a$ includes a derivative by $x_1$ before setting $x_1=0$, so we do not expect an equivalent combination for the vacuum seagull, and indeed $x_1=0$ is not a singularity of its equation set. All the other loci are also loci of the vacuum seagull, and therefore we will find that their solutions are only composed of $E_1-E_5,E_2,E_3,E_4$ and have analogues for the vacuum seagull.

Using (\ref{sing_loc_sol}) we find:
\be\label{x1 solution}
I^{PS}\Big|_{x_1=0}=-\frac{(O_2-O_5)I^{PS}}{x_{25}}\Big|_{x_1=0}~.
\ee
This result can be gotten by setting $x_1=p^2=0$ in the original integral\footnote{setting $p^2=0$ is equivalent to taking $p^\mu=0$ in the case of a self-energy diagram.} (\ref{ps_int}), which leaves the resulting vacuum integral:
\bea
&&\int \frac{d^dl_1\, d^dl_2}{(l_1^2-x_2)(l_2^2-x_3)((l_1+l_2)^2-x_4)(l_1^2-x_5)}\\
&&=\frac{1}{x_{25}}\Big[\int \frac{d^dl_1\, d^dl_2}{(l_1^2-x_2)(l_2^2-x_3)((l_1+l_2)^2-x_4)}-\int \frac{d^dl_1\, d^dl_2}{(l_2^2-x_3)((l_1+l_2)^2-x_4)(l_1^2-x_5)}\Big]\nonumber\\
&&=-\frac{(O_2-O_5)I^{PS}}{x_{25}}\Big|_{x_1=0}\nonumber.
\eea

By multiplying \eqref{ps_set k} with $k_a$ we get a generalization of this result
\begin{equation}
\begin{split}
    x_{25}(d-2)I_k^{PS}\Big|_{x_1=0}=&\Big[(d-2)(O_5-O_2)+k O_2\Big]I_k^{PS}\Big|_{x_1=0}\\&-k\big[x_{34}(2O_5-3O_2)+(2x_2+O_2)(O_3-O_4)\Big]I_{k-1}^{PS}\Big|_{x_1=0}
\end{split}
\end{equation}
This expresses the numerator integrals $I_k^{PS}$, at the singular locus, in terms of descendant diagrams and it generalizes the solution for the numerator-free integral.

\subsubsection{$x_5=0$}

In this case we get the solution
\bea\label{x5 locus}
I^{PS}\Big|_{x_5=0}&=&\frac{(O_2-O_5)I^{PS}}{x_{12}}\\
&&+\frac{1}{d-2}\Big[\frac{2x_3\partial_3 -2x_4\partial_4}{x_{34}}(O_3+O_4-O_5)I^{PS}\Big|_{x_5=0} -\frac{2x_1\partial_1-2x_2\partial_2}{x_{12}}O_5I^{PS}\Big]_{x_5=0} ~.\nonumber
\eea
Using equations for the sources in appendix \ref{app: sources} we reproduce the result in equation (6) of \cite{Kniehl:2016yrh}, namely
\bea
I^{PS}\Big|_{x_5=0}&=& \Big[\frac{O_2I^{PS}}{x_{12}}-\frac{(O_3-O_4)I^{PS}}{x_{34}}\\&&-\frac{x_{12}(2x_3\partial_3 -2x_4\partial_4 )-x_{34}(4x_2 \partial_2 + 2x_3 \partial_3 + 2x_4 \partial_4 + 8-3d)}{(d-2)x_{12}x_{34}}O_5 I^{PS} \Big]_{x_5=0}
 \nonumber
~.
\eea

As expected we also have a corresponding solution for the vacuum seagull which is identical to (\ref{x5 locus}) up to an $O_1$ term
\bea 
I^{VS}\Big|_{x_5=0}&=& \Big[-\frac{(O_1-O_2)I^{VS}}{x_{12}}-\frac{(O_3-O_4)I^{VS}}{x_{34}}\\&&-\frac{x_{12}(2x_3\partial_3 -2x_4\partial_4 )-x_{34}(4x_2 \partial_2 + 2x_3 \partial_3 + 2x_4 \partial_4 + 8-3d)}{(d-2)x_{12}x_{34}}O_5 I^{VS} \Big]_{x_5=0}
 \nonumber
~.
\eea

\subsubsection{$\lambda_A=0$}
The solution on this locus is
\be
I^{PS}\Big|_{\lambda_A=0}= \Big[ \frac{(d-2)(O_2-O_5)I^{PS}}{2(d-3)s_A^2}-\frac{s_A^2\partial_1+s_A^1\partial_2}{(d-3)x_5} O_5I^{PS} \Big]_{\lambda_A=0}.
\ee
As before we can bring this expression into a form that will be identical to the vacuum seagull solution for this locus up to an $O_1$ term:

\begin{equation}
I^{PS}\Big|_{\lambda_A=0}=\frac{1}{d-3}\Big[\frac{(d-2)O_2 I^{PS} }{2s_A^2}+\frac{(s_A^5+x_2)\partial_2+x_3\partial_3 +x_4\partial_4 +\frac{1}{2}(8-3d)}{s_A^2}O_5 I^{PS} \Big]_{\lambda_A=0},
\end{equation}

\bea
I^{VS}\Big|_{\lambda_A=0} &=& \frac{1}{d-3}\Big[\frac{(d-2)O_2 I^{VS} }{2s_A^2} +\frac{(d-2)O_1 I^{VS} }{2s_A^1} \non
&& +\frac{(s_A^5+x_2)\partial_2+x_3\partial_3 +x_4\partial_4 +\frac{1}{2}(8-3d)}{s_A^2}O_5 I^{VS} \Big]_{\lambda_A=0} ~.
\eea

The generalization to higher $k$ is the recursion relations
\begin{equation}
        (d-3+k)s_A^2 I_k\Big|_{\lambda_A=0}-k\,x_{34}(x_1+s_A^5)I_{k-1}\Big|_{\lambda_A=0}=J^{PS/VS}_k\Big|_{\lambda_A=0},
\end{equation}
with
\begin{equation}
    \begin{split}
       J^{PS}_k\Big|_{\lambda_A=0}=& \Big[[(s_A^5+x_2)\partial_2+x_3\partial_3+x_4\partial_4+\frac{1}{2}(8-2k-3d)]O_5+\frac{d+k-2}{2}O_2\Big]I_k^{PS}\Big|_{\lambda_A=0}\\&+k[\frac{1}{2}x_{34}(2O_5-3O_2)+((x_2+s_A)+\frac{1}{2}(O_3-O_4)O_2)]I^{PS}_{k-1}\Big|_{\lambda_A=0}
    \end{split}
\end{equation}
and
\begin{equation}
\begin{split}
    J^{VS}_k\Big|_{\lambda_A=0}=&\Big[[(s_A^5+x_2)\partial_2+x_3\partial_3+x_4\partial_4+\frac{1}{2}(8-2k-3d)]O_5\\&+\frac{d+k-2}{2}(O_2+\frac{s_A^2}{s_A^1}O_1)\Big]I_k^{VS}\Big|_{\lambda_A=0}+k(x_1+s_A^5)(O_3-O_4)I_{k-1}^{VS}\Big|_{\lambda_A=0}
    \end{split}
\end{equation}

This is a first order recursion relation for numerator integrals $I_k^{VS/PS}$, valid at the singular locus. It generalizes the solution for the numerator-free integral, and it is goes beyond the general recursion relation \eqref{recursion equation} which is second order.

\subsubsection{$\lambda_B=0$}
On this locus the solution happens to be identical for both the propagator and vacuum seagull and is
\be
I^{PS/VS}\Big|_{\lambda_B=0}=\frac{s_B^4\partial_3 +s_B^3\partial_4}{x_5(d-3)}(O_3+O_4-O_5)I^{PS/VS} \Big|_{\lambda_B=0}.
\ee

\subsection{3 scale sector for the vacuum seagull with numerator} 
\label{3-scale_sector}

In this subsection, we specialize to a 3 mass scale sector where we are able to obtain a closed-form solution to the numerator integral $I_1^{VS}$. We choose this sector to be $x_2 = x_4 =0$ where both the sources and $I^{VS}$ are known in closed-form. At this sector the numerator equation (\ref{numerator equation}) becomes 
\be \label{single_numerator_vacuum_seagull}
x_5(d-2)I_{1}^{VS}=(d-2)x_{1}\,x_{3}I^{VS}-J_1^{VS} ~.
\ee
and $J_1^{VS}$ and $I^{VS}$ are known in closed form \cite{Burda:2017tcu}.

\subsubsection{Solution for general $d$}

The vacuum seagull 3-mass scale sector solution for general $d$ is
{\small \bea
I^{VS}(x_1,0,x_3,0,x_5) & =& i\pi^{\frac{3d}{2}} (x_5)^{\frac{3d}{2}-5} \,\big((1-x)(1-y)\big)^{d-3} \non
&& \Big\{ \Big[-3\Gamma\big(3-\frac{3d}{2}\big)\Gamma(4-d)\Gamma\big(\frac{d}{2}-2\big)\Gamma\big(\frac{d}{2}-1\big)\, x\,y^{\frac{3d}{2}-5} {_{2}F_{1}}\big(5-\frac{3d}{2},4-d,3-\frac{d}{2} ; \frac{x}{y}\big) \label{i1} \non
&&- \frac{4\pi \mathrm{Csc}(\frac{\pi d}{2})\Gamma(2-d)\Gamma\big(2-\frac{d}{2}\big)}{3d-8}\,x^{\frac{d}{2}-1}y^{d-3}{_{2}F_{1}}\big(3-d,2-\frac{d}{2},\frac{d}{2}-1 ; \frac{x}{y}\big)\Big]\non
&& \times F_1(3d/2-4,d-2,d-3,3d/2-3 ; \, x,y) \non
&& + x \leftrightarrow y \non
&& + \pi\, d\,\mathrm{Csc}\big(\frac{\pi d}{2}\big)\Gamma(2-d)\Gamma\big(-\frac{d}{2}\big)\, x^{\frac{d}{2}-1}  {_{2}F_{1}}(d/2-1,d-2,d/2 ;\,x) + x \leftrightarrow y\non
&& +\Gamma\big(1-\frac{d}{2}\big)^3\, (x\,y)^{\frac{d}{2}-1}{_{2}F_{1}}(d/2-1,d-2,d/2; \,x) \,{_{2}F_{1}}(d/2-1,d-2,d/2; \,y)\non
&&+\frac{2\pi^3\,\mathrm{Csc}^2\big(\frac{\pi d}{2}\big)\mathrm{Csc}\big(\frac{3\pi d}{2}\big)\Gamma\big(\frac{d}{2}-1\big)}{(d-2)(\Gamma(d-2))^2} \Big\}
\eea}
where $x=x_1/x_5$ and $y=x_3/x_5$.

In this sector the source $J_1$, given in simplified form in equation (\ref{num source}), has the form 
\bea \label{J1_sector}
J_1^{VS} &=& -x_3 (O_1 I^{VS} - O_2 I^{VS}) - ( d -2 ) x_1 (O_3I^{VS} - O_4I^{VS}) + (8 - 
    3 d) x_3 O_5I^{VS} \non
    &&  + 2 x_3 (-x_1 + x_3) \partial_3 O_5I^{VS} -\frac{1}{2}(x_3-x_5)^2(\partial_3 O_1 I^{VS}- \partial_4O_1 I^{VS}-\partial_3 O_2I^{VS} + \partial_4O_2I^{VS}) \non
&&+ 1/2 [(x_3 + x_5) (\partial_3 O_4 O_1I^{VS} - \partial_3 O_5 O_1I^{VS} ) + (x_3 - 
       x_5) (\partial_4 O_3 O_1I^{VS} - \partial_4 O_5 O_1I^{VS})]  \non    
 &&- 1/2 [(x_3 + x_5) (\partial_3 O_4 O_2I^{VS} - \partial_3 O_5 O_2I^{VS} ) + (x_3 - 
       x_5) (\partial_4 O_3 O_2I^{VS} - \partial_4 O_5 O_2I^{VS})] . \non
\eea
Each of the sources can be computed in terms of the simpler diagrams $\mathrm{Dia}_{n_1,n_2,n_3}(x_1,x_2,0)$, and $\mathrm{Tad}_n(x)$.
\begin{subequations} \label{sources_of_J1}
\bea 
O_1 I^{VS} &=& \mathrm{Tad}_1(0) \mathrm{Dia}_{1, 1, 1}( x_3, x_5, 0)  \\
O_2 I^{VS} &=& \mathrm{Tad}_1(x_1) \mathrm{Dia}_{1, 1, 1}( x_3, x_5, 0) \\
O_3 I^{VS}  &=& \mathrm{Tad}_1(0) \mathrm{Dia}_{1, 1, 1}( x_1, x_5, 0) \\
O_4 I^{VS} &=& \mathrm{Tad}_1(x_3) \mathrm{Dia}_{1, 1, 1}( x_1, x_5,0)\\
O_5 I^{VS} &=& \mathrm{Melon}_{1,1,1,1}(x_1,x_3,0,0)= G(1, 1) \mathrm{Dia}_{1, 1, 2 - d/2}( x_1, x_3, 0)\\
\partial_3 O_5 I^{VS} &=& \mathrm{Melon}_{1,2,1,1}(x_1,x_3,0,0)=  G(1,1) \mathrm{Dia}_{1,2, 2-d/2}(x_1,x_3,0)\\
\partial_3 O_1I^{VS} &=& \mathrm{Tad}_1(0)\mathrm{Dia}_{2, 1, 1}( x_3, x_5, 0)  \\
\partial_4 O_1I^{VS} &=& \mathrm{Tad}_1(0) \mathrm{Dia}_{1, 1, 2}( x_3, x_5, 0) \\
\partial_3 O_2I^{VS} &=& \mathrm{Tad}_1(x_1)\mathrm{Dia}_{2, 1, 1}( x_3, x_5, 0) \\
\partial_4 O_2I^{VS} &=& \mathrm{Tad}_1(x_1) \mathrm{Dia}_{1, 1, 2}( x_3, x_5, 0)\\
\partial_3O_4O_1I^{VS} &=&  \mathrm{Tad}_2(x_3) \mathrm{Tad}_1(0) \mathrm{Tad}_1(x_5)  \\
\partial_3 O_5 O_1I^{VS} &=& \mathrm{Tad}_2(x_3) \mathrm{Tad}_1(0) \mathrm{Tad}_1(0)  \\
\partial_4 O_3 O_1I^{VS} &=& \mathrm{Tad}_1(0)\mathrm{Tad}_1(x_5)\mathrm{Tad}_2(0) \\
\partial_4 O_5 O_1I^{VS} &=& \mathrm{Tad}_1(0)\mathrm{Tad}_1(x_3)\mathrm{Tad}_2(0) \\
\partial_3O_4O_2I^{VS} &=&  \mathrm{Tad}_2(x_3) \mathrm{Tad}_1( x_1) \mathrm{Tad}_1(x_5)  \\
\partial_3 O_5 O_2I^{VS} &=& \mathrm{Tad}_2(x_3) \mathrm{Tad}_1( x_1) \mathrm{Tad}_1(0) \\
\partial_4 O_3 O_2I^{VS} &=& \mathrm{Tad}_1(x_1)\mathrm{Tad}_1(x_5)\mathrm{Tad}_2(0)\\
\partial_4 O_5 O_2I^{VS} &=& \mathrm{Tad}_1(x_1)\mathrm{Tad}_1(x_3)\mathrm{Tad}_2(0) 
\eea
\end{subequations}
where
\be
G(n_1,n_2)= i^{1-d}\pi^{d/2} \frac{\Gamma(n_1+n_2-d/2)\Gamma(d/2-n_1)\Gamma(d/2-n_2)}{\Gamma(n_1)\Gamma(n_2)\Gamma(d-n_1-n_2)},
\ee

\be
\mathrm{Tad}_n(x)=i^{1-d}\pi^{d/2}\frac{\Gamma(n-d/2)}{\Gamma(n)}(-x)^{d/2-n},
\ee
and 
{\small\bea
&&\mathrm{Dia}_{n_1,n_2,n_3}(x,y,0)=i^{2-2d} \pi^{d}(-y)^{d-n_1-n_2-n_3}\frac{\Gamma(d/2-n_3)}{\Gamma(n_1)\Gamma(n_2)\Gamma(n_3)\Gamma(d/2)} \non& &\Big[ \Gamma(\frac{d}{2}-n_1)\Gamma(n_1+n_2+n_3-d)\Gamma(n_1+n_3-\frac{d}{2}){_{2}F_{1}}\Big( n_1+n_2+n_3-d,n_1+n_3-\frac{d}{2},n_1-\frac{d}{2}+1;\frac{x}{y}\Big)\non
& &+\Big(\frac{x}{y}\Big)^{d/2-n_1} \Gamma(n_1-\frac{d}{2}) \Gamma(n_2+n_3-\frac{d}{2})\Gamma(n_3){_{2}F_{1}}\Big(n_2+n_3-\frac{d}{2},n_3,\frac{d}{2}-n_1+1;\frac{x}{y}\Big)\Big].
\eea}
For the definitions of $\mathrm{Tad}_n(x), \mathrm{Dia}_{n_1,n_2,n_3}(x,y,z)$ and $\mathrm{Melon}_{n_1,n_2,n_3,n_4}(x,y,z,u)$ in terms of Feynman integrals see Appendix \ref{sources_feynman_integrals}. Note that in dimensional regularization we can set $\mathrm{Tad}_1(0)=\mathrm{Tad}_2(0)=0$.

We have checked the resulting expression for $J_1^{VS}$ by computing it through (\ref{vs_w_numerator_source}) and comparing the two results numerically at random points in parameter space and dimension $d$, to find full agreement. 

\subsubsection{Epsilon expansion of the 3 scale sector of the vacuum seagull} \label{3-scale_sector_epsilon}
In this section we use the epsilon expansion of the sources given in the previous section, to obtain the epsilon expansion of $I_1(x_1,0,x_3,0,x_5)$.  We follow the notations, conventions and results of \cite{Martin:2016bgz, Martin:2017lqn}. According to their conventions for $d=4-2\epsilon$ each loop should be multiplied by a ``loop factor" $C = (16\pi^2) \frac{\mu^{2\epsilon}}{(2\pi)^d}$, where $\mu$ is a regularization mass scale. There is also a relative sign difference between the conventions in this paper and the conventions of  \cite{Martin:2016bgz, Martin:2017lqn} for the diagrams ``$\mathrm{Tad}$", ``$\mathrm{Dia}$" and ``$I^{VS}$".  We will omit overall phases.

The $\epsilon$-expansions of the diagrams which appear in the source terms (\ref{sources_of_J1}), are \cite{Martin:2016bgz, Martin:2017lqn} 
{\small\bea
\mathrm{Tad}_1(x) &=& \frac{1}{\epsilon}x - A(x) - \epsilon A_\epsilon(x)\\
\mathrm{Dia}_{1,1,1}(x,y,z) &=& \frac{1}{2 \epsilon^2} (x+y+z) - \frac{1}{\epsilon} \left(A(x) + A(y) + A(z) -\frac{1}{2}(x+y+z) \right) -I_0(x,y,z) \non \\
\mathrm{Melon}_{1,1,1,1}(x,y,z,u) &=& \frac{1}{3 \epsilon^3}(xy + xz + xu + yz + yu + zu) \non
&& +\frac{1}{\epsilon^2} \Big[-\frac{1}{2}((y+z+u)A(x)+(x+z+u)A(y) \non
&&+(x+y+u)A(z)+(x+y+z)A(u)) +\frac{1}{3}(xy + xz + xu + yz + yu + zu)\non
&&-\frac{1}{12}(x^2+y^2+z^2+u^2) \Big] \non
&&+ \frac{1}{\epsilon} \Big[A(x)A(y)+A(x)A(z)+A(x)A(u)+A(y)A(z)+A(y)A(u)+A(z)A(u)\non
&&-(y+z+u)(A_\epsilon(x)+A(x))/2 - cyclic \non
&& +(x A(x)+ y A(y)+ z A(z) + u A(u))/4 +\frac{1}{3}(xy + xz + xu + yz + yu + zu) \non
&& -3 (x^2+y^2+z^2+u^2)/8\Big].
\eea}
Here $A(x) = x \left(\ln(x/Q^2)-1 \right)$ and $A_\epsilon(x) = x \left(-\frac{1}{2} \ln^2(x/Q^2)+\ln(x/Q^2)-1 -\pi^2/12 \right)$, where $Q^2 = 4\pi e^{-\gamma_E}\mu^2$ and $\gamma_E$ is Euler's gamma.  For $I_0(x,y,z)$ all we need is the case $I_0(0,y,z)$ (see \cite{Martin:2016bgz})
\bea \label{I0xyz}
    I_0(0,y,z) &=& A_\epsilon(y)+A_\epsilon(z) + (y-z)\Big[\mathrm{Li}_2(1-y/z)+\frac{1}{2}\ln^2(z/Q^2)\Big] -y \ln (y/Q^2)\ln(z/Q^2) \non
    &&+2 y \ln (y/Q^2) +2z \ln (z/Q^2) -\frac{5}{2}(y+z)\Big].
\eea

The sources $\mathrm{Tad}_2(x)$, $\mathrm{Dia}_{2,1,1}(x,y,z)$ and $\mathrm{Melon}_{2,1,1,1}(x,y,z,u)$ are essentially derivatives with respect to the relevant masses:
\bea
\mathrm{Tad}_2(x) &=& \frac{\partial}{\partial x}\mathrm{Tad}_1(x) = \frac{1}{\epsilon} - \ln (x/Q^2) + \epsilon \left(\frac{\pi^2}{12} + \frac{1}{2} \ln^2(x/Q^2) \right)\\
\mathrm{Dia}_{2,1,1}(x,y,z) &=& \frac{\partial}{\partial x} \mathrm{Dia}_{1,1,1}(x,y,z)= \frac{1}{y-z} \Big[z \frac{\partial}{\partial z} \mathrm{Dia}_{1,1,1}(x,y,z) -y \frac{\partial}{\partial y} \mathrm{Dia}_{1,1,1}(x,y,z)\non
&&- \Tad_2(x)(\Tad_1(z) - \Tad_1(y)) \Big] \\
\mathrm{Melon}_{2,1,1,1}(x,y,z,u) &=& \frac{\partial}{\partial x} \mathrm{Melon}_{1,1,1,1}(x,y,z,u) = \frac{1}{3 \epsilon^3}(y + z + u) +\dots 
\eea
We provided two equivalent expressions for $\mathrm{Dia}_{2,1,1}(x,y,z)$, based on SFI equations for the diameter diagram \cite{Kol:2018kga}, so that we can use (\ref{I0xyz}) which is more convenient than the general expression for $I_0(x,y,z)$.

Plugging these expressions into (\ref{J1_sector}) with the explicit expressions given in (\ref{sources_of_J1}), we find 
\be
J_1^{VS} = \frac{1}{\epsilon^3} J_1^{(-3)} + \frac{1}{\epsilon^2} J_1^{(-2)} + \frac{1}{\epsilon} J_1^{(-1)} + O(\epsilon^0),
\ee
where
{\small\bea
J_1^{(-3)} &=& \frac{1}{3} x_1 x_3 (x_1+x_3) \\
J_1^{(-2)} &=& \frac{1}{3} [x_1 x_3 (5 (x_1 + x_3) + 6 x_5(1-\ln(x_5/Q^2))  \non
&& + 3 (-x_1 + x_5) \ln(x_1/Q^2) + 3 (-x_3 + x_5) \ln(x_3/Q^2) )] \\
J_1^{(-1)} &=& \frac{1}{12}x_1 x_3 [ (76+\pi^2)(x_1+x_3) +96 x_5(1-\ln(x_5/Q^2)) \non && +6(x_1-x_5)\ln^2(x_1/Q^2)+6(x_3-x_5)\ln^2(x_3/Q^2) \non
   && -24 x_5 \ln(x_1/Q^2)\ln(x_3/Q^2) -12 (x_1+x_3-3x_5)\ln^2(x_5/Q^2) \non
   && +12 \ln(x_1/Q^2)(-5x_1 +x_5 +2x_1 \ln (x_5/Q^2) ) +12 \ln(x_3/Q^2)(-5x_3 +x_5 +2x_3 \ln (x_5/Q^2) ) \non
   && +24 (-x_1+x_5) \mathrm{Li}_2 (1-x_1/x_5) +24 (-x_3+x_5) \mathrm{Li}_2 (1-x_3/x_5) ]~.
\eea}
We have confirmed this result by testing each term in the expansion against the result (\ref{J1_sector}) at random points in the 3-mass sector, getting numerically very close to $d=4$.  This was done by setting very high numerical precision. 

The other ingredient for computing the $\epsilon$-expansion of the vacuum seagull with numerator in the 3-mass sector through Equation (\ref{single_numerator_vacuum_seagull}) is the $\epsilon$-expansion for the vacuum seagull in this sector, which can be found in \cite{Burda:2017tcu}, and is repeated here for completeness:
\be
I^{VS} = \frac{1}{\epsilon^3} I^{(-3)} + \frac{1}{\epsilon^2} I^{(-2)} + \frac{1}{\epsilon} I^{(-1)} + O(\epsilon^0),
\ee
where
\bea
I^{(-3)} &=& \frac{1}{6} [x_1+x_3 +2x_5] \\
I^{(-2)} &=& x_1(1-1/2 \ln (x_1/Q^2)) + x_3(1-1/2\ln (x_3/Q^2)) +x_5(5/3-\ln (x_5/Q^2))] \\
I^{(-2)} &=& -\frac{2}{3}x_5 -\frac{2}{3}(x_1+x_3 +2x_5)\ln(1-x_3/x_1)\ln(1-x_5/x_1) \non
&& + 3\big(x_1(1-1/2 \ln x_1) + x_3(1-1/2\ln x_3) +x_5(5/3-\ln (x_5/Q^2))\big)(2-\ln (x_5/Q^2)) \non
&& + \frac{x_5}{4}( x_1/x_5 \ln^2(x_1/x_5) + x_3/x_5 \ln^2(x_3/x_5))  \non
&& + \frac{1}{6}(x_1+x_3 +2x_5)\big(7 +\pi^2/4 -9/2(2-\ln (x_5/Q^2))^2\big) \non
&& +  x_5(1-x_1/x_5)\mathrm{Li}_2(1-x_1/x_5) + x_5 (1-x_3/x_5)\mathrm{Li}_2(1-x_3/x_5)\non
&& + \frac{2}{3} (x_1+x_3 +2x_5) \big( \mathrm{Li}_{1,1}(x_1/x_3, x_3/x_5) + \mathrm{Li}_{1,2}(x_3/x_1, x_1/x_5)~.
\eea
We are now in a position to write the $\epsilon$-expansion for the vacuum seagull with numerator
\be
I_1^{VS} = \frac{1}{\epsilon^3} I_1^{(-3)} + \frac{1}{\epsilon^2} I_1^{(-2)} + \frac{1}{\epsilon} I_1^{(-1)} + O(\epsilon^0),
\ee
with
\bea
I_1^{(-3)} &=& \frac{1}{x_5}\big(x_1 x_3 I^{(-3)} -\frac{1}{2} J_1^{(-3)} \big)\\
I_1^{(-2)} &=& \frac{1}{x_5} \big( x_1 x_3 I^{(-2)} -\frac{1}{2}(J_1^{(-3)}+J_1^{(-2)})  \big)\\
I_1^{(-1)} &=& \frac{1}{x_5} \big( x_1 x_3 I^{(-1)} - \frac{1}{2} (J_1^{(-3)}+J_1^{(-2)}+J_1^{(-3)}) \big) ~.
\eea

\subsection{Homogeneous solutions}
\label{subsec:homog}

In this subsection we discuss two kinds of homogeneous solutions to the xSFI equations. We note that the maximal cut is known to be a homogeneous solution of the differential equations \cite{Lee:2012te, Remiddi:2016gno, Primo:2016ebd}. 

\subsubsection{Homogeneous solution for the propagator seagull}
One important result of the SFI method is the representation of a Feynman diagram as a line integral over the sources. Such a solution is obtained by solving the homogeneous set of SFI equations and then using variation of parameters to find the full solution.  The first step is then to find the homogeneous solution.  

For the vacuum seagull the homogeneous solution was determined in \cite{Burda:2017tcu}, and for the propagator seagull it was determined up to a free function $g(\phi)$ in \cite{Kol:2018ujm}.  Given the new equation for the propagator seagull in (\ref{ps_set}) the homogeneous solution becomes unique, and is given by
\be
I_0^{PS}(x)=(-x_1)^{1-\frac{d}{2}} (x_5)^{1-\frac{d}{2}}(\lambda_A \, \lambda_B)^{\frac{d-3}{2}}.
\ee
This is the same as the homogeneous solution for the vacuum seagull apart for the simple pre-factor $(-x_1)^{1-\frac{d}{2}}$.

\subsubsection{Solving for the generating function}

Here we solve the differential equation for the generating function of numerator integrals \eqref{recursion diff form}. By setting the r.h.s.  to zero we can solve for the homogeneous solution
\be
G_{0}=z^{3-d}(x_5-2x_{12}x_{34}z+B z^{2})^{\frac{d-4}{2}}~.
\ee
We can also set only the source terms to zero, equivalent to making the recursion relation homogeneous, and the solution of
the resulting equation is

\bea
\frac{1}{I}g &=& x_5(x_5-(x_{12}x_{34}-\sqrt{x_{12}^2 x_{34}^2-x_5B})z)^{-1} \\ &\,&_{2}F_{1}\left(1,\frac{d-2}{2};d-2;\frac{2\sqrt{x_{12}^2 x_{34}^2-x_5B}z}{x_5-(x_{12}x_{34}-\sqrt{x_{12}^2 x_{34}^2-x_5B})z}\right)
\eea
which will generate $I_{n}$ up to source terms 
\be
I_{n}=\frac{g^{(n)}}{n!}\mid_{z=0}+ \textrm{sources}~.
\ee

\section{Extended SFI equations and the SFI group} \label{Sec:xSFI_general}

In this section we discuss extended SFI equations for a general diagram. 

\presub {\bf Extended SFI variations}. Consider a general diagram with $L$ loops and $X$ external legs. Let us denote a basis for loop currents by $l_a, ~ a=1, \dots, L$, a basis for external currents $p_u, ~u=1, \dots, X-1$ and the union of the two bases by $\{q_r\}=(l_1, \dots, l_L, p_1, \dots, \, p_{X-1})$.

SFI variations, or pre-equations, were defined in \cite{Kol:2015gsa,Kol:2016hak,Kol:2016veg,Burda:2017tcu} by  \bea
 \delta l_a &=& \ep_a^{~r}\, q_r \non
 \delta p_u &=& \ep_u^v \, p_v
\label{SFI_var}
 \eea
The variations in the first line are the same as those of IBP, and are known to define the Lee Lie group \cite{LeeGroup2008}, while those in the second line are the same as DE.

Extended SFI considers the following larger set of variations \bea
 \delta l_a &=& \ep_{aR}^{~r}\, (q \cdot q)^R \, q_r \non
 \delta p_u &=& \ep_{uU}^v \, (p \cdot p)^U\, p_v
\label{XSFI_var}
 \eea
where $R=\{ n_{rs} \}$ is a multi-index so that $(q \cdot q)^R= \prod_{r,s}  \, (q_r \cdot q_s)^{n_{rs}}$ and similarly for $U$. This is a non-linear extension of \eqref{SFI_var} which preserves Lorentz invariance. Geometrically, it describes vector fields on the space of currents. 

The extended SFI variations define a graded Lie algebra \be
 H = \bigoplus_{k=0,1,2,\dots} H_k
 \ee
 where the grading is defined by $k=|R|=\sum n_{rs}$. $H_0$ includes the Lee Lie group.

\presub {\bf Induced action on diagram parameters}. The $H_0$ variations within \eqref{XSFI_var} induce an action (a representation) on the quadratics \be
 Q = \mathrm{Sp} \{ q_r \cdot q_s \} 
\label{def:Q}
\ee
where $Sp$ denotes the span. 

Recall that SFI \cite{Kol:2016hak} gives a special role to the subspace of squares within $Q$, $S \subset Q$ defined by \be
S=  \mathrm{Sp} \{ k_i^2 \} \oplus \mathrm{Sp} \{ p_u \cdot p_v \}
\label{def:S}
\ee
where $k_i$ are edge or propagator currents (considered as a linear combination of the $q_r$). The SFI group $G \subset H_0$ is defined to be the sub-algebra which preserves $S$ (as a subspace rather than pointwise). In particular, the representation of $H_0$ on $Q$ restricts to \be
 \mbox{a representation of $G$ on $S$.}
 \ee

Going beyond SFI, the space of irreducible numerators $N$ is defined to be the quotient space \be
 N := Q/S
 \label{def:N}
 \ee
By the defining property of $G$, its $Q$ representation also defines
\be
 \mbox{a representation of $G$ on $N$.}
 \ee

In the presence of numerators, the most general Feynman integral associated with a given diagram or graph is \be
I( \{ m_i^2 \}, \{p_u \cdot p_v\},K) = \int dl \frac{N^K}{\prod_i (k_i^2-m_i^2)}
\ee
where $dl$ denotes the integration over all loop currents, $K=(k_1,\dots, k_{Num})$ is a multi-index so that $N^K:=\prod_i N_i^{k_i}$ where $N_1,\dots,N_{Num}$ is a basis of irreducible numerators. The parameter space $\hat{X}$ for $I$ consists of mass-squares and kinematical invariants, together with the multi-index K.

As in SFI, a variation in $H$ generates a partial differential equation for $I$ (with a recursion on $K$). More precisely, the action of $H$ on $Q$ determines this equation. The homogeneous part of the equation defines a vector field on parameter space $\hat{X}$ (together with raising and lowering operators on $K$). This creates \be
\mbox{an action, or representation, of $H$ on $\hat{X}$.}
\ee 

\presub {\bf Action of SFI group on the extended SFI equations}. Here we study the algebraic structure of the extended SFI equation system.

An SFI equation has the form \be
 (T^a)_i^j \, x_i \, \del^j \, I = \dots
\ee
where $a$ is the equation number, $x_i$ denotes variables in parameter space $X$ (masses and kinematical invariants), $\del^j \equiv \del/\del x_j$, $I$ is the integral under study, $(T^a)_i^j$ are $x$-independent constants and the ellipsis denote the source terms, which are independent of $I$. The equation defines a differential operator acting on $X$  \be
 g^a=(T^a)_i^j \, x_i \, \del^j
\ee
The commutation relations among the operators $g^a$ define the SFI Lie algebra $G$. 

Schematically we may write $g = x \del$. An extended SFI equation has the schematic form \be
 p(x,z,1/z) \, \del\,  I = \dots
\ee
where $z$ denotes formal parameters, each associated with an irreducible numerator, $p(x,z,1/z)$ is a polynomial, and $\del$ denotes both $\del/\del x$ and $\del/\del z$.  The equation defines an operator \be
 u =  p(x,z,1/z) \, \del
 \label{def:u}
 \ee
If $p=p(x)$ is linear in $x$ and independent of $z$ then $u \in G$.  The operator subspace spanned by linear combinations of $u$'s (with $x$-independent coefficients) is denoted by \be
M := {\rm Sp} \{u^a \} \supset G.
\ee

Given any two operators $u, v \in M$ of the form \eqref{def:u} their commutation relation would be of the form \be
[u,v] = \left[p_m(x)\, \del, q_n(x)\, \del \right] = r_{m+n-1}(x)\, \del 
\ee
where $m,n$ are the corresponding mass-squared dimensions of the polynomials. It can be seen that the result is an operator that is still first order in derivatives. However, the mass-squared dimension of $r(x)$ increases as long as $m,n > 1$.  Therefore generically $[u,v] \notin M$ and hence $M$ is not a Lie algebra. 

This means that the extended SFI equation system does not define a Lie algebra anymore. However, for all $g \in G,\, u \in M$ \be
 [g,u] \in M
 \ee
 and hence, by definition, \emph{$M$ is a module over the SFI group $G$} (equivalently, $M$ is a representation of $G$). In this way, the SFI group $G$ is extended to a module $M$ over $G$, while $G$ continues to play a central role. This conclusion holds for the seagull diagrams studied in this paper 
 as well as more generally for any diagram.

In SFI, the $X$ space was foliated into orbits of $G$. The extension of this property is a foliation into the invariant manifolds of $M$. 

\section{Summary and discussion} 
\label{summary}

For a given diagram topology (or graph), SFI considers the associated Feynman Integral to be a function of its most general parameters, and it strives to formulate a complete equation system generated by current variations. So far, these parameters included the masses and the kinematical invariants. In this paper, we added another class of parameters, namely numerator powers, within the concrete context of the seagull diagrams, which have a single irreducible numerator. In order to extend the equation system, we extended the current variations under consideration. 

Our main results are of two kinds: new equations and new evaluations. We find two new equations \begin{itemize}
    \item Eq. \eqref{recursion equation} is a second order recursion relation in the numerator power. It applies to both seagull diagrams, only the form of the sources is different (\ref{seagull_w_numerator_source},\ref{vs_w_numerator_source}). This equation is transformed into an equivalent differential equation \eqref{recursion diff form} acting on a generating function.
    \item The second row in the equation system (\ref{ps_set}) extends the SFI equation system for the numerator-free propagator seagull. Even though it is numerator-free it can be generated only by an extended SFI variation of the schematic form $\del_l \cdot q (q \cdot q)$.
\end{itemize}
The main novel evaluation extends the evaluation of the numerator-free vacuum seagull within a 3 mass scale sector \cite{Burda:2017tcu} to $I_1^{VS}$, which includes the numerator to the first power:  \begin{itemize}
    \item Eq. \eqref{single_numerator_vacuum_seagull} presents $I_1^{VS}$ in terms of $I^{VS},\, J_1^{VS}$, which are detailed below it. The relevant $\epsilon$-expansions around $d=4$ are detailed in subsection \ref{3-scale_sector_epsilon}. 
\end{itemize}
In addition, we present the solutions in the revised singular locus in subsection \ref{singular loci}, and some homogeneous solutions in subsection \ref{subsec:homog} (the maximal cut is known to be one).

In conclusion, this paper provides the first study of numerator integrals within the (extended) SFI method, and the first closed-form evaluation of the above-mentioned numerator seagull.

\presub {\bf Open questions}. While there are infinitely many variations of the form $\del_q \cdot q (q \cdot q)^n$ the associated equations are found to be all generated by a finite number of equations. It would be interesting to study this property for general diagrams. 

Another open question concerns the generating function of numerator seagulls $G(\{x\};z)$ \eqref{def:Gz}. It is a function of 6 variables: 5 $x$'s and $z$, the formal parameter. The tetrahedron is a diagram which resolves the quartic vertex of the vacuum seagull, and it has 6 variables. It would be interesting to study the relation of $G(\{x\};z)$ with the tetrahedron $I(x)$.  

\subsection*{Acknowledgments}

It is a pleasure to thank K. Larsen for discussions.
This work was supported in part by the ``Quantum Universe" I-CORE program of the Israeli Planning and Budgeting Committee.

\appendix

\section{Source equations} \label{app: sources}
We have used SFI equations for sources of the seagull equation in order to simplify some expressions. In (\ref{ps_set}) we used
\begin{equation}
\partial_{3}O_4 I^{PS}_1=(-x_{12}+O_2)O_4\partial_3 I^{PS},
\end{equation}
\begin{equation}
\partial_{4}O_3 I^{PS}_1=(x_{12}-O_2)O_3\partial_4 I^{PS},
\end{equation}
\begin{equation}
(\partial_3-\partial_4)O_5 I^{PS}_1=(4-2d-4x_1 \partial_1+4x_2\partial_2+(O_2-x_{12})(\partial_3+\partial_4))O_5I^{PS},
\end{equation}
\begin{equation}
(s_B^4\partial_3+s_B^3\partial_4+2x_5\partial_5)O_2 I^{PS}=(d-3+\frac{1}{2}(\partial_3+\partial_4)(O_3+O_4-O5))O_2I^{PS}.
\end{equation}

These equations can be generalized to any numerator power giving us the simplifications used in (\ref{ps_set k})
\begin{equation}
\partial_3O_4I_{k+1}=(-x_{12}+O_2)O_4\partial_3 I_k+2k(2x_1+2x_2-x_5+2O_2-O_5)O_4I_{k-1},
\end{equation}
\begin{equation}
\partial_4O_3I_{k+1}=(x_{12}-O_2)O_3\partial_4 I_k+2k(2x_1+2x_2-x_5+2O_2-O_5)O_3I_{k-1},
\end{equation}
\begin{equation}
\begin{split}
(\partial_3-\partial_4)O_5 I^{PS}_{k+1}=&(4-2d-4x_1 \partial_1+4x_2\partial_2+(O_2-x_{12})(\partial_3+\partial_4))O_5I_k^{PS}\\&+2k(x_{34}+O_3-O_4)I_{k-1},
\end{split}
\end{equation}
\begin{equation}
\begin{split}
(s_B^4\partial_3+s_B^3\partial_4+2x_5\partial_5)O_2 I^{PS}_k=&(d-3+\frac{1}{2}(\partial_3+\partial_4)(O_3+O_4-O5))O_2I^{PS}_k\\&-k(x_{34}+O_3-O_4)O_2I_{k-1}.
\end{split}
\end{equation}

In order to simplify (\ref{numerator equation}) we used some of the above equations as well as
\begin{equation}
(s_B^4\partial_3+s_B^3\partial_4)O_5I_1^{PS}=(x_{34}(8-3d+4x_2\partial_2+2x_3\partial_3+2x_4\partial_4)+(O_2-x_{12})(s_B^4\partial_3-s_B^3\partial_4))O_5I^{PS}.
\end{equation}

\section{Integral forms} \label{sources_feynman_integrals}
The integral definitions of the diagrams used in section \ref{3-scale_sector} are
\begin{align}
    \mathrm{Tad}_n(x)&=\int \frac{d^dl}{(l^2-x)^n},\\
    \mathrm{Dia}_{n_1,n_2,n_3}(x,y,z)&=\int \frac{d^dl_1 d^dl_2}{(l_1^2-x)^{n_1}(l_2^2-y)^{n_2}((l_1+l_2)^2-z)^{n_3}},\\
 \mathrm{Melon}_{n_1,n_2,n_3,n_4}(x,y,z,u)&=\int \frac{d^dl_1\,d^dl_2\,d^dl_3}{(l_3^2-x)^{n_1}((l_1+l_3)^2-y)^{n_2}(l_2^2-z)^{n_3}((l_1+l_2)^2-u)^{n_4}}.
\end{align}


\end{document}